# *Vortex Lattice in a Rotating Bose-Einstein Condensate*

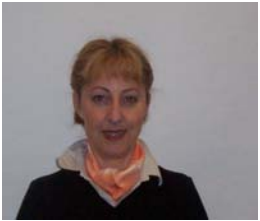

*By: Enikő Madarassy*

A Bose-Einstein condensate *(BEC)* is a state of matter of a system of bosons confined in an external potential. The atoms are cooled to temperatures very near to absolute zero. Under this condition quantum effects become evident on a macroscopic scale. This follows from the fact that a large fraction of the atoms collapse into the lowest quantum state of the external potential. The history of BEC has its origin in Satyendra Nath Bose's [1] and Albert Einstein's [2] works in the 1920s on the quantum statistics of particles with integer spin.

The first weakly-interacting atomic Bose-Einstein condensate was produced by Eric Cornell and Carl Wieman in 1995, at the University of Colorado at Boulder NIST-JILA lab. They used a gas of rubidium atoms cooled to 170 nanokelvin (nK). A dilute, weakly-interacting BEC was also obtained in a confined ultracold gas of sodium and lithium atoms with the help of cooling and trapping techniques [3]. Cornell, Wieman and Wolfgang Ketterle were awarded the 2001 Nobel Prize in Physics for this achievement.

One can create a cloud of dilute BEC by confining in a magneto-optical trap about $10^9$ atoms with the help of laser beams and magnetic fields. In that state, the temperature is $T \approx 10^{-5} K$ and the number density is $n \approx 10^{10}\ cm^{-3}$. To obtain BEC, the temperature of the cloud is reduced ($T \approx 10^{-6}\ K$ and $n \approx 10^{14}\ cm^{-3}$) by radio-frequency pulses to evaporate higher energy atoms in magnetic trap with a harmonic confinement [4]. The particle interactions are low-energy, two-body collisions described by the atomic s-wave scattering length, *a*. More than *99%* of the atoms are condensed at zero temperature. The interactions are weak due to the scattering cross-section is much less than the mean space between particles.

Dilute BECs, regulated and managed by electromagnetic and optical means, have been produced with Lithium , Sodium , spin-polarised Hydrogen, meta-stable Helium, Potassium , Caesium , Ytterbium ... Combination of experimental and theoretical informations of BEC give us a good insight in fundamental concepts of condensed matter physics.

## *Vortices*

The most striking properties of superfluids are the creation and observation of quantum vortices. Solving an equation, which describes the bosons at very low temperatures, the Gross-Pitaevskii equation *(GPE),* we can see that it allows solutions, which are topologically non-trivial e.g. vortices with zero density and non-zero fluid circulation.

A vortex is manifested as a density hole within the condensate. In superfluids, vortices are characterised by quantised circulations. They can decay under collision at the boundary of the condensate or due to other dissipative mechanism.

The conditions for vortex creation depend upon the shape of the condensate and the form of the trapping potential. The standard method of generating
quantised vortices in a superfluid is by rotation about a fixed axis when quantised vortex lines appear aligned with the axis of rotation. At low rotation frequencies the superfluid remains stationary. Provided that the rotation is greater than a critical value, one or more vortices form at the edge and enter the condensate. Their presence reduces the free energy of the system, and they become energetically favourable, above a certain critical rotation frequency [5].

## *Trapped and Dissipative Condensate Harmonic Trap*

The achievement and studies of BECs require cooling of metastable atomic samples. In a trapped gas, the BEC can be regarded as a coherent standing matter wave. The harmonic confinement of the 2D BECs is described by the trapping potential, $V_{tr}$.

Let us begin with the dimensional geometry in 3D, which has a cylindrical symmetry about the z-axis:

$$V_{tr}(z,r) = \frac{m}{2}(\omega_z^2 z^2 + \omega_r^2 r^2).$$

Under a transverse confinement, the dynamics of the system become quasi-2D and the vortex line becomes rectilinear. In this system, the 2D GPE yields a good description of the condensate.

In our simulation, we use a 2D dimensionless form:

$$V_{tr}(r) = \frac{1}{2}(\omega_r^2 r^2),$$

where, $r = \sqrt{(1+\varepsilon_x)x^2 + (1+\varepsilon_y)y^2}$, with $\varepsilon_x = 0.03$, $\varepsilon_y = 0.09$ and $\omega_r = 2\pi \times 219\,Hz$, corresponding to the *ENS experiment*, see [6].

Here, $\varepsilon_x$ and $\varepsilon_y$, describe small deviations of the trap from the axi-symmetry. So, the condensate is elongated along the *x-axis* due to this small anisotropy and the boundary surface of the condensate becomes unstable. This phenomenon plays a key role in the formation of vortices.

## *Dissipative Regime*

To understand several experiments with vortices and solitons, it is helpful to include the dynamical coupling of the condensate to the thermal cloud, the effect of dimensionality and the role of quantum fluctuations.

In superfluid helium the problem of quantised vortices and mutual friction force on the vortex were well described at finite temperature [7], when on account of the presence of a thermal cloud, dissipation arises. This non-condensed part acts as a source of dissipation, having as a result, the damping of excitations like collective modes.

The mean-field and collisional damping with uniform densities were well understood. Mean field coupling between the condensate and thermal cloud leads to a lower frequency oscillation. In that case for example, a soliton loses energy to the thermal cloud [8].

The decay of a vortex at finite temperatures was investigated with the help of the GPE and a Boltzmann kinetic equation for the thermal cloud [9]. In a trapped Bose gas, the two-fluid hydrodynamics of the condensate and non-condensate was described by including the dissipation, associated with viscosity and thermal conduction.

Inclusion of a damping term, $\gamma$ into the GPE [10], models dissipative losses that occur in real environments. Collective damped oscillations was noticed due to some dissipative mechanism [11].

We can state, that the dynamical dissipation factor, $\gamma$, plays a significant role in condensates, which in our equation:

$$(i-\gamma)\hbar\frac{\partial \Psi(r,t)}{\partial t} = \left[-\frac{\hbar^2}{2m}\nabla^2 + V_{tr}(r,t) + g|\Psi(r,t)|^2 - \mu - \Omega L_z\right]\Psi(r,t)$$

is the $\gamma$ - term.

## *Energies*

Introduction of the $\gamma$ parameter changes the nature of the *GPE*, so the total energy now is not conserved. The vortex is thermodynamically unstable [12]. At the centre of the trap, a vortex has maximum energy, because of the locally homogeneous density. In order to perform our analysis, we decompose the total energy, $E_{tot}$, into kinetic, internal, quantum and trap contributions,

$$E_{tot} = E_{kin} + E_{int} + E_q + E_{trap},$$

where,

$$E_{kin}(t) = \int \frac{\hbar^2}{2m} \left( \sqrt{\rho(x,t)}\, v(x,t) \right)^2 d^2r,$$

$$E_{int}(t) = \int g(\rho(x,t)) d^2r,$$

$$E_q(t) = \int \frac{\hbar^2}{2m} \left( \nabla \sqrt{\rho(x,t)} \right)^2 d^2r,$$

and

$$L_z = i\hbar \left( y \frac{\partial}{\partial x} - x \frac{\partial}{\partial y} \right).$$

Notice that, the kinetic energy, ($E_{kin}$) is related with the velocity field. The internal energy, ($E_{int}$) represents the internal energy of the fluid. The quantum energy, ($E_q$) comes from the gradient of the condensate and the trap energy, ($E_{trap}$) has its origin from the applied trap potential.

## *An Array of Vortices*

Let us study first the variation of different energies and the *z-component* of the angular momentum with time and with different dissipations for an array of vortices, (see Fig.(1-9)).

So, for $\Omega = 0.75$, the vortices come in automatically in the condensate. To have the right shape and initial conditions (the Thomas-Fermi approximation), the *GPE* first runs in imaginary time. Then the program is switched to real time with suitable values for $\Omega$ and $\gamma$.

We have a considerable change in the values of different energies and $L_z$ in that time, when the vortices come in in the condensate. (see Fig. (1-6), for *C=1400*). Here, C is the dimensionless form of the coupling constant, g. ( $g = \frac{4\pi\hbar^2 a}{m}, C = \frac{4\pi a}{L}$, L is the size of the system along the z-axis, a is the scattering length and m is the mass of the atoms).

For a vortex lattice, when $\Omega = 0.75$, the increasing rate of $E_{tot}, E_{kin}, E_{trap}, E_q$ and $L_z$ and the decay rate of $E_{int}$ as a function of dissipations: $\gamma = 0.003, 0.03$ and $0.07$, was studied by Fig. (7-9).

## *Vortex Lattice Formation in a Rotating Bose-Einstein Condensate*

We solve numerically the Gross-Pitaevskii equation (*GPE*). First, we have an equilibrium condensate (with stationary potential, $\Omega = 0$), with a circular shape, (see Fig. 10).

The dynamics of the condensate density, $|\Psi|^2$ is investigated with the help of some contour-plots of the density (the left plots), together with some plots of the phase (the right plots) see Fig. (10-16). This method is a direct proof of a test of our computer code.

As a consequence of setting for $\Omega = 0.85$, the trapping potential begins to rotate. Because of the small anisotropy of the trapping potential, $V_{tr}$, the condensate is elongated, (see Fig.11 and Fig. 12).

Later, since the boundary surface is unstable, the surface with low curvature becomes host to surface waves, (see Fig. 12). These ripples make themselves visible at vortex cores, (see Fig.13 and Fig. 14).

More and more vortices enter the condensate, (see Fig. 15) and form a vortex lattice, (see Fig. 16). At that point, the angular momentum is transferred into quantised vortices and the condensate recovers its circular like form.

## *Figures:*

*For $\Omega = 0.75$, the log (base e) of different energies and $L_z$ as a function of time, compared with $f(x) = a + bx$, for that time, when the vortices come in in the condensate.*

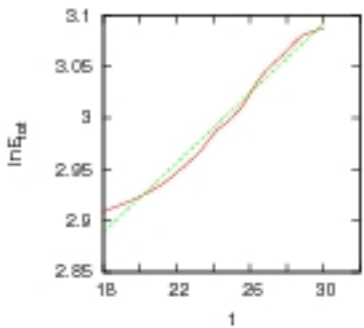

*Fig.1: Total energy,
　　　a=2.5832, b=0.01696*

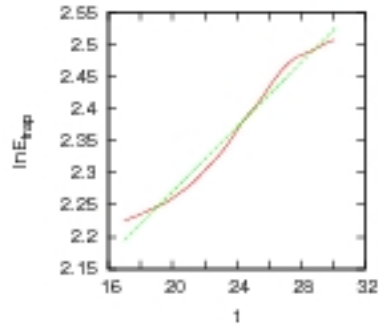

*Fig.2: Trap energy,
　　　a=1.7665, b=0.02519*

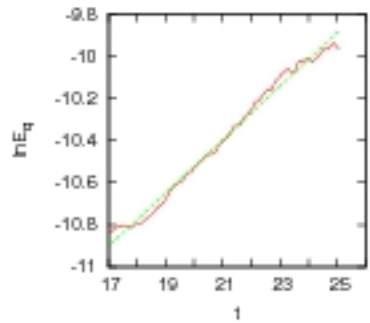

*Fig.3: Quantum energy,
　　　a=-13.0613, b=0.12690*

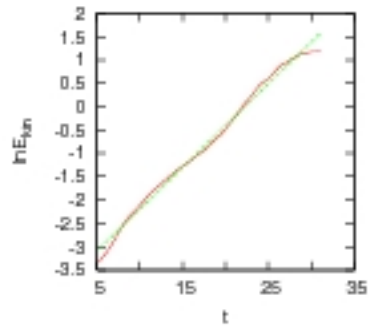

*Fig.4: Kinetic energy,
　　　a=-3.9922, b=0.17922*

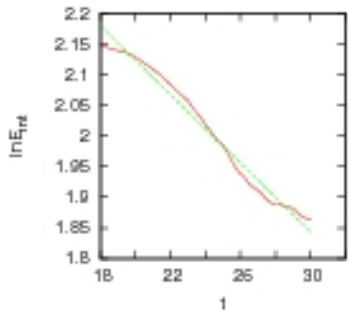

*Fig.5: Internal energy,*
*a=2.6821, b=-0.02793*

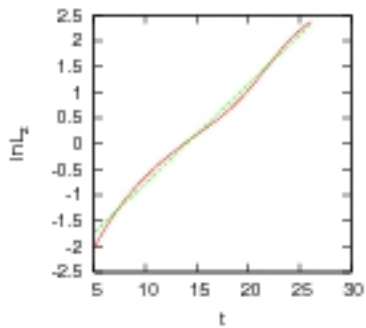

*Fig.6: z-component of the angular*
*momentum, a=2.7013, b=0.19277*

$L_z$ = *the z-component of the angular momentum*

$\Omega$ = *the angular velocity of the trap*

$\gamma$ = *the dissipation*

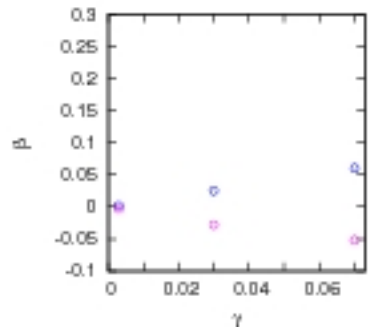

*Fig. 7: The increasing rate of the trap energy (blue circles) and the decay rate of the internal energy (purple circles) as a function of $\gamma$.*

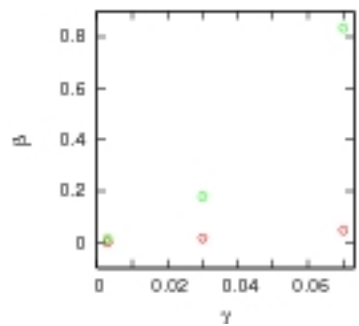

*Fig. 8: The increasing rate of the total energy (red circles) and kinetic energy (green circles) as a function of $\gamma$.*

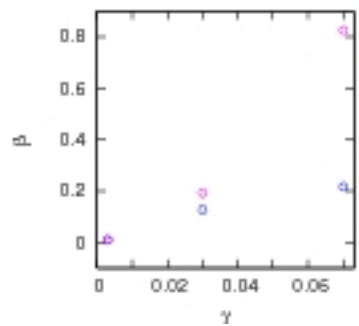

*Fig. 9: The increasing rate of the quantum energy (blue circles) and $L_z$ (purple circles) as a function of $\gamma$.*

*Fig. 12: t = 13.0*

*Figures:*

*For Ω = 0.85,*
*Left: Density profile and*
*Right: Phase profile of the condensate at:*

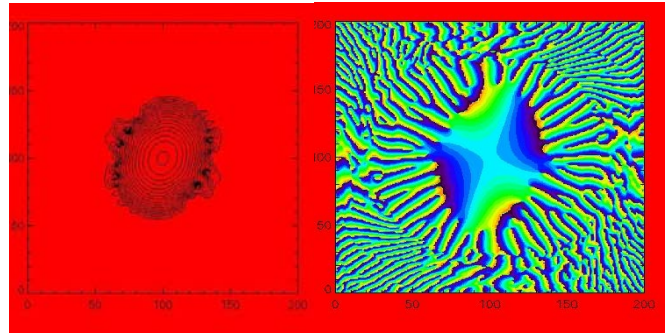

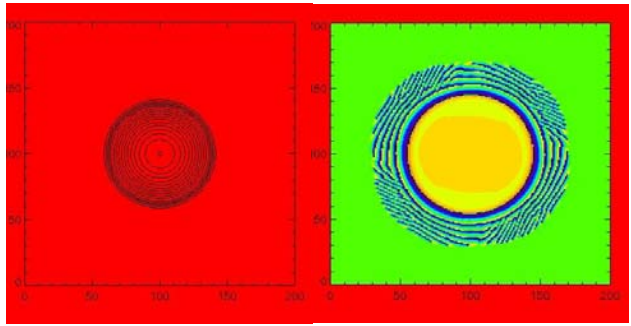

*Fig. 10: t = 0.2*

*Fig. 13: t = 16.4*

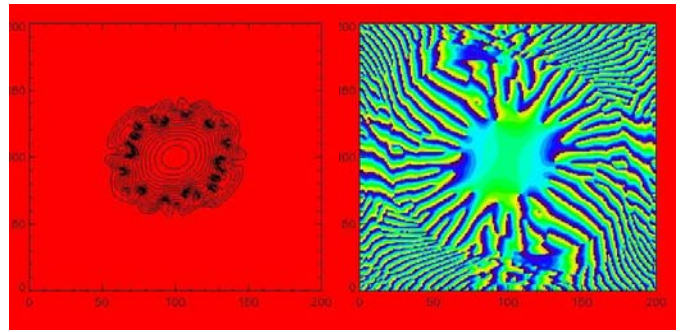

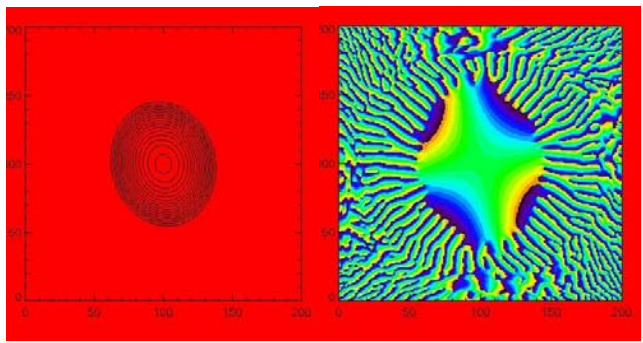

*Fig. 14: t = 22.4*

*Fig. 11: t = 11.0*

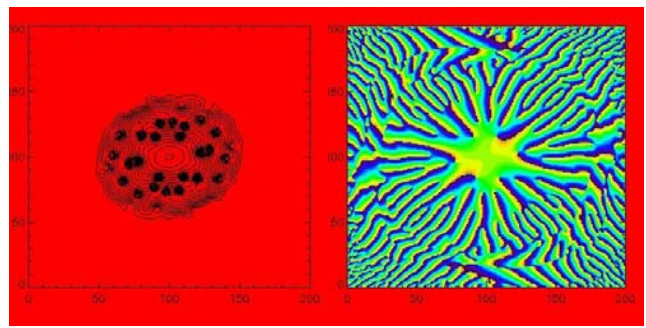

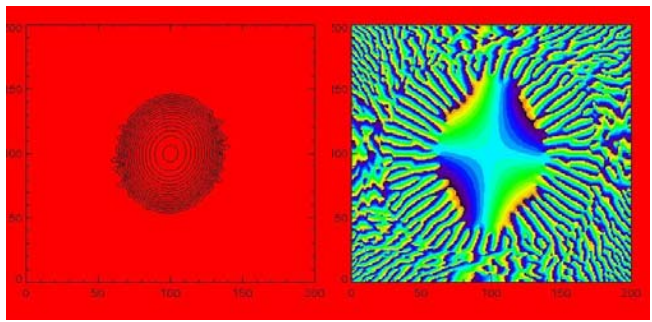

*Fig. 15: t = 26.8*

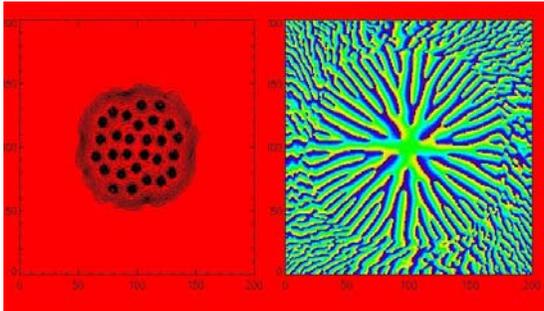

*Fig. 16: t = 54.8*